\documentclass{aa}  

\usepackage{graphicx}
\graphicspath{{figs/}}
\newcommand{\ol}{\overline}
\usepackage{txfonts}
\usepackage{bm}
\usepackage[colorlinks,allcolors=blue]{hyperref}

\begin{document}

   \title{Effects of the centrifugal force in stellar dynamo simulations}

   \author{Felipe H. Navarrete\inst{1}\fnmsep\thanks{e-mail: felipe.navarrete@hs.uni-hamburg.de}
          \and
          Petri J. K\"apyl\"a\inst{2,3} 
          \and
          Dominik R.G. Schleicher\inst{4}
          \and
          Robi Banerjee\inst{1}
          }

   \institute{Hamburger Sternwarte, Universit\"at Hamburg, Gojenbergsweg
       112, 21029 Hamburg, Germany\\
             \email{felipe.navarrete@hs.uni-hamburg.de}
         \and
             Leibniz-Institut f\"ur Sonnenphysik (KIS), Sch\"oneckstr. 6, 79104
             Freiburg, Germany
         \and
             Institut f\"ur Astrophysik und Geophysik, Georg-August-Universit\"at
             G\"ottingen, Friedrich-Hund-Platz 1, 37077, G\"ottingen, Germany
         \and
             Departamento de Astronom\'ia, Facultad de Ciencias F\'isicas y
             Matem\'aticas, Universidad de Concepci\'on, Av. Estebar Iturra s/n
             Barrio Universitario, Casilla 160-C, Chile
             }

   \date{Received --; accepted --}

 
  \abstract
  {The centrifugal force is often omitted from simulations of stellar convection
   either for numerical reasons or because it is assumed to be weak compared to the gravitational force. However, the centrifugal force might be an important factor in rapidly rotating stars,
   such as solar analogs, due to its $\Omega^2$ scaling, where
   $\Omega$ is the rotation rate of the star.}
   {We study the effects of the centrifugal force in a set of 21 semi-global
   stellar dynamo simulations with varying rotation rates. Included in the set are three control runs aimed at distinguishing the effects of
   the centrifugal force from the nonlinear evolution of the
   solutions.}
   {We solved the 3D magnetohydrodynamic equations with the
   {\sc Pencil Code} in a solar-like convective zone in a spherical
   wedge setup with a $2\pi$ azimuthal extent. The rotation rate
   and the amplitude of the centrifugal force were varied. We
   decomposed
   the magnetic field into spherical harmonics and studied the migration of azimuthal
   dynamo waves (ADWs), the energy of different large-scale magnetic
   modes, and differential rotation.}
   {In the regime with the lowest rotation rates, $\Omega = 5-10\Omega_\odot$, where $\Omega_\odot$
   is the rotation rate of the Sun, we see no marked changes in
   either the differential
   rotation or the magnetic field properties. For intermediate
   rotation,
   $\Omega = 20-25\Omega_\odot$, we identify an increase in the
   differential
   rotation as a function of centrifugal force. The axisymmetric magnetic energy
   tends to decrease with centrifugal force, while the non-axisymmetric
   one increases.
   The ADWs are also affected, especially in the propagation direction.
   In the most rapidly rotating set with $\Omega=30\Omega_\odot$, these changes are more
   pronounced, and in one case the propagation direction of the ADW changes
   from prograde to retrograde. The control runs suggest that the results are
   a consequence of the centrifugal force and not due to the details of the initial
   conditions or the history of the run.}
   {We find that the differential rotation and properties of the ADWs
only     change as a
   function of the centrifugal force when rotation is rapid enough.}


   \maketitle

\section{Introduction}

Simulations of stellar convection, usually aimed at explaining solar
phenomena, often omit the centrifugal force. This is due to the assumption that
its amplitude is small due to the relatively slow rotation of the Sun.
Earlier in its history, however, the Sun must have been rotating much more 
rapidly because, in general, stars are born with larger angular momenta that are slowly reduced via magnetic braking \citep{Skumanich72,Matt12}.
Therefore, the influence
of the centrifugal force is expected to be more important at earlier
stages because
its amplitude increases as the square of the rotation rate. To study these phases
of rapid ration in the solar context, such as magnetic field evolution,
one has to study young solar analogs at earlier phases that are rotating much faster than the Sun \citep[e.g.,][]{Lehtinen16}. This allows us to
study the evolution of the Sun up to the present, given that outflows that are
produced by magnetic braking do not significantly affect the structure of the
star, only the rotation rate.

Observations by 
\citet{Lehtinen16} of magnetic fields of solar analogs reveal that they
are active and show a characteristic split between
the axisymmetric and non-axisymmetric spot distributions. These authors also estimate that, among
solar-like stars with non-axisymmetric spot distribution, the active longitude
periods are shorter than the rotation period of the star.
One plausible explanation is the presence of azimuthal dynamo waves (ADWs). These
waves propagate in the rotating frame of reference of the star either
in prograde or retrograde fashion with a uniform frequency
irrespective of the underlying fluid motions. Such solutions were first
discovered in linear mean-field dynamo models
\citep[e.g.,][]{Krause80}. To explain their observations,
\cite{Lehtinen16} argue that the propagation of the ADWs must be
prograde.

V530 Per is an extreme case of a rapidly rotating Sun-like star with an
estimated rotation period of $0.32$~days \citep{Cang20}, which corresponds to
about $75\Omega_\odot$, where $\Omega_\odot$ is the rotation rate of the Sun. 
This makes the gravitational force at its surface only 9.5 times larger than
the centrifugal force. In comparison, in the Sun this ratio is
$5.3\times10^4$. There are also clear differences between the magnetic field of
V530~Per and the Sun. For example, \cite{Cang20} also find that the magnetic
field distribution of V530~Per is asymmetric with respect to the equator.
It is characterized by a stronger magnetic field near the north
pole, with a peak field strength of 1~kG. It is as yet unclear why similar stars
have different field strengths and symmetries, but there are indications that
rotation may play an important role in the magnetic activity of Sun-like stars
\citep{Lehtinen16} as well as low-mass stars \citep[e.g.,][]{Reiners22}.

Such rapid rotation is commonly found in close binaries
if tidal locking is assumed. For example, V471~Tau is a post-common-envelope
binary in which the secondary is a main-sequence solar-like star with a
mass
of $0.93M_\odot$, a radius of $0.96R_\odot$, and a binary period of about
0.5 days \citep[e.g.,][]{Voelschow16}. If tidal locking is assumed, this gives a
ratio of gravitation to centrifugal forces of about 22. Interestingly,
\cite{Zaire22} analyzed the magnetic activity of the K2 star in V471 Tau
and find that the magnetic field is also dominated by a concentration
on one hemisphere. They also find that the spot coverage and
brightness map, derived from Zeeman-Doppler imaging,
do not follow the magnetic activity cycle inferred from H$\alpha$
variability. This suggests that it might be inappropriate to use spot coverage to study magnetic cycles in rapidly
rotating stars \citep{Zaire22}.

Simulations of stellar dynamos often produce ADWs whose characteristics change
with the rotation rate and the physics involved. \cite{Cole14} studied the
propagation properties of ADWs in a set of three runs with moderate rotation
rates of up to $6.7$ times the solar value with 3D magnetohydrodynamic
simulations. They find that the waves have a rotation rate that is slower than that of
the gas, that is, they are retrograde. The magnetic structure in the
ADW propagates like a rigid body, and therefore such motion cannot be
explained by advection by the fluid in a differentially rotating
convection zone. This result was
later confirmed by \cite{Viviani18} with a larger set of runs. Most of their
runs show retrograde ADWs independently of the rotation rate, but in some cases
standing or prograde waves appeared. Recently, \cite{Viviani21}
presented a set of four runs with moderate rotation rates where the usual
prescribed radial dependence of the radiative heat conductivity was replaced
by the more realistic Kramers opacity law \citep{Brandenburg00,2017ApJ...845L..23K}.
This suggests that using this heat conductivity might affect the
direction of the propagation of ADWs indirectly by affecting the flow
through the pressure gradient and/or dissipation. However, it
might come with the
cost of pushing the transition point of differential rotation profiles of
simulations from anti-solar profiles to solar-like profiles to even larger Coriolis numbers \citep{Viviani21}.

\cite{Navarrete22b} explored the effect of the centrifugal force in
the context of changes in the internal structure of the stars, with
the aim to check whether the resulting changes are sufficient to explain the
observed eclipsing time variations in post-common-envelope binaries, as proposed
in the Applegate scenario \citep{Applegate92}.
In this paper we study the effects of centrifugal force in semi-global dynamo
simulations further. We focus on differential rotation, magnetic energy, and ADW
propagation. In Sect.~\ref{sec:model} we present the model and the
implementation of the centrifugal force. Section~\ref{sec:results} presents
the results, and our conclusions are drawn in Sect.~\ref{sec:conclusions}.
\section{Model}\label{sec:model}

We solved the fully compressible magnetohydrodynamic equations in a spherical
grid with coordinates $(r,\Theta,\phi),$ where $0.7R \leqslant r \leqslant R$ is
radius and $R$ is the radius of the star, $\pi/12\leqslant \Theta \leqslant
11\pi/12$ is the colatitude, and $0\leqslant \phi < 2\pi$ is the longitude. The
model is the same as in \citet{Kapyla13} and \citet{Navarrete20, Navarrete22}.
The equations adopt the following forms:
\begin{align}
        \frac{\partial\vec{A}}{\partial t} &= \vec{u}\times\vec{B} -
        \eta\mu_0\vec{J}, \\
        \frac{{\rm D}\ln \rho}{{\rm D}t} &= -{\bm\nabla}{\bm\cdot} \vec{u},\\
        \frac{{\rm D}\vec{u}}{{\rm D}t} &= \vec{\mathcal{F}}^{\rm grav} +
        \vec{\mathcal{F}}^{\rm Cor} + \vec{\mathcal{F}}^{\rm cent} -
        \frac{1}{\rho}(\bm\nabla p - \vec{J}\times\vec{B} - \bm\nabla\bm\cdot2\nu\rho
        \vec{S}), \\
        T\frac{{\rm D}s}{{\rm D}t} &= \frac{1}{\rho}\left[\eta\mu_0\vec{J}^2 -
        \bm\nabla\bm\cdot(\vec{F}^{\rm rad} + \vec{F}^{\rm SGS})\right] + 
        2\nu\vec{S}^2, \label{eq:entropy}
\end{align}
where $\vec A$ is the magnetic vector potential, ${\vec B} = {\vec\nabla}\times
{\vec A}$ is the magnetic field, $\vec u$ is the velocity field, $\eta$ is the
magnetic diffusivity, $\mu_0$ is the vacuum permeability, $t$ is the time,
$\vec J = \vec \nabla \times \vec B / \mu_0$ is the electric current density,
$\rho$ is the mass density, $p$ is the pressure, $\nu$ is the viscosity,\begin{equation}
    S_{ij} = \frac{1}{2}(u_{i;j} - u_{j;i})-\frac{1}{3}\delta_{ij}\vec\nabla\bm\cdot\vec u
\end{equation}
is the rate-of-strain tensor, where semicolons
denote covariant differentiation, $T$ is the temperature, and $s$ is the
specific entropy.
Furthermore, $\vec F^{\rm rad} = -K\nabla T$ is the radiative flux,
which we modeled with the diffusion approximation, where $K=K(r)$ has
a fixed spatial profile (see Sect.~2.1 in \citealt{Kapyla14}). We also
investigated the effects of 
Kramers opacity in some runs (see Sect.~\ref{sec:adw}).
We did this by replacing
the radiative heat conductivity, $K,$ in the radiative flux term
$\vec{F}^{\rm rad} = -K\bm\nabla T$ with\begin{equation}\label{eq:Kramers}
    K=K_0\left(\frac{\rho}{\rho_0}\right)^{-(a+1)}\left(\frac{T}{T_0}\right)^{3-b},
\end{equation}
where $a=1$ and $b=-7/2$ correspond to the Kramers opacity law
\citep[][]{Brandenburg00}. Here, $K_0$ is a
constant that depends on natural constants and, in simulations, on
the luminosity of the model \citep{Viviani21}. The $\vec F^{\rm SGS}=-\chi_{\rm SGS}\rho T\vec\nabla s$ is a sub-grid scale
flux that we implemented to smooth grid-scale fluctuations  that would otherwise
make the system unstable. Here, $\chi_{\rm SGS}$ is the sub-grid scale entropy
diffusivity, and it varies smoothly from $0$ at $r/R = 0.7$ to $0.4\nu$ at
$r/R=0.72$; it then smoothly increases by a factor of $12.5$ at $r/R=0.98$,
above which it is constant. The first three terms on the right-hand side of
Eq.~(\ref{eq:entropy}),
\begin{align}
       & \vec{\mathcal{F}}^{\rm grav} = -(GM/r^2)\hat{\vec{r}},\\
       & \vec{\mathcal{F}}^{\rm Cor}  = -2\vec{\Omega}_0\times\vec{u},\\
       & \vec{\mathcal{F}}^{\rm cent} = -c_f\vec{\Omega}_0\times(\vec{\Omega}_0\label{eq:cforce}
        \times\vec{r}),
\end{align}
are the gravitational, Coriolis, and centrifugal forces.

\subsection{Boundary and initial conditions}
The magnetic field follows a perfect conductor condition at the bottom
of the convective zone and
is radial at the surface. The temperature gradient was kept fixed at the
bottom, whereas at the top we applied a black-body condition. For the entropy and
density, we assumed a vanishing first derivative at both latitudinal boundaries.
The latitudinal boundaries are stress-free and perfectly conducting.
The initial state is isentropic. Perturbations were introduced by
initializing the magnetic and velocity fields with low-amplitude
Gaussian white noise.

\subsection{Centrifugal force}
The parameter $c_f$ in Eq.~(\ref{eq:cforce}) was
introduced by \citet{Kapyla20b} and controls the strength of the centrifugal
force. A $c_f$ value of $1$ corresponds to the unaltered centrifugal force amplitude, and
$c_f = 0$ implies no centrifugal force. It is defined as
\begin{equation}
        c_f = \frac{\left|\vec{\mathcal{F}}^{\rm cent}\right|}
        {\left|\vec{\mathcal{F}}^{\rm cent}_0\right|},
\end{equation}
with $|\vec{\mathcal{F}}^{\rm cent}_0|$ being the unaltered magnitude of the
centrifugal force. The necessity of controlling the centrifugal force is due to
the enhanced luminosity and rotation rate in simulations of compressible stellar
(magneto-)convection. Similarly to \citet{Kapyla20b}, each run was initialized
with $c_f=0$ and was increased in small incremental steps after the
saturated regime is reached.

To get a sense of how strong the centrifugal force is in our simulations,
we computed the ratio
of centrifugal to gravitational forces in the simulations as well as in a
real Sun-like star with the same rotation rate. We defined the ratio between
the two as 
\begin{equation}
        \mathcal{F} = \frac
                       {(|\vec{\mathcal{F}}^{\rm cent}|/
                       |\vec{\mathcal{F}}^{\rm grav}|)_{\rm sim}}
                       {(|\vec{\mathcal{F}}^{\rm cent}|/
                       |\vec{\mathcal{F}}^{\rm grav}|)_\star},
\end{equation}
where the subscripts $\star$ denote the real star and ``sim'' the simulations.
These values are shown in the last column of Table.~\ref{tab:runs}.
By using a value as low as $c_f=10^{-4}$, our simulations are influenced by
the centrifugal force just below the value that the equivalent star with the same
rotation rate and radius would have,
and the simulations that have the strongest centrifugal force have
$\mathcal F = 87$.

\begin{table*}
\caption{Summary of  the simulation parameters.}
\label{tab:runs}
\centering
\begin{tabular}{c c c c c c c c c c}
  \hline\hline
  Run & $\Omega/\Omega_\odot$ & $c_f$ & Co & Ta & Re & Rm
      & $\langle\Delta_\Omega^{(60^\circ)}\rangle_t$ 
      &  $\langle\Delta_\Omega^{(r)}\rangle_t$
      & $\mathcal{F}$
      \\
  \hline
C1 & 5  & 0         & $7.3$ & $1.6\times10^{8}$ & $3.0\times10^{3}$ & $3.0\times10^{3}$ & $4.2\times10^{-2}$ &  $2.5\times10^{-2}$ & 0 \\
C2 & 5  & $10^{-4}$ & $7.3$ & $1.6\times10^{8}$ & $3.0\times10^{3}$ & $3.0\times10^{3}$ & $4.3\times10^{-2}$ &  $2.5\times10^{-2}$ & 0.87 \\
C3 & 5  & $10^{-3}$ & $7.3$ & $1.6\times10^{8}$ & $3.0\times10^{3}$ & $3.0\times10^{3}$ & $4.4\times10^{-2}$ &  $2.6\times10^{-2}$ & 8.7 \\
C4 & 5  & $10^{-2}$ & $6.8$ & $1.6\times10^{8}$ & $3.2\times10^{3}$ & $3.2\times10^{3}$ & $4.8\times10^{-2}$ &  $2.8\times10^{-2}$ & 87 \\
D1 & 10 & 0         & $20$  & $6.3\times10^{8}$ & $2.2\times10^{3}$ & $2.2\times10^{3}$ & $1.3\times10^{-2}$ &  $5.6\times10^{-3}$ & 0 \\
D2 & 10 & $10^{-4}$ & $20$  & $6.3\times10^{8}$ & $2.2\times10^{3}$ & $2.2\times10^{3}$ & $1.4\times10^{-2}$ &  $6.1\times10^{-3}$ & 0.87 \\
D3 & 10 & $10^{-3}$ & $20$  & $6.3\times10^{8}$ & $2.2\times10^{3}$ & $2.2\times10^{3}$ & $1.4\times10^{-2}$ &  $6.2\times10^{-3}$ & 8.7  \\
D4 & 10 & $10^{-2}$ & $20$  & $6.3\times10^{8}$ & $2.2\times10^{3}$ & $2.2\times10^{3}$ & $1.5\times10^{-2}$ &  $7.0\times10^{-3}$ & 87  \\
E1 & 20 & 0         & $57$  & $2.5\times10^{9}$ & $1.6\times10^{3}$ & $1.6\times10^{3}$ & $3.4\times10^{-3}$ &  $9.0\times10^{-4}$ & 0 \\
E2 & 20 & $10^{-4}$ & $56$  & $2.5\times10^{9}$ & $1.6\times10^{3}$ & $1.6\times10^{3}$ & $3.7\times10^{-3}$ &  $9.9\times10^{-4}$ & 0.87 \\
E3 & 20 & $10^{-4}$ & $54$  & $2.5\times10^{9}$ & $1.6\times10^{3}$ & $1.6\times10^{3}$ & $3.7\times10^{-3}$ &  $8.1\times10^{-4}$ & 0.87 \\
E4 & 20 & $10^{-3}$ & $53$  & $2.5\times10^{9}$ & $1.7\times10^{3}$ & $1.7\times10^{3}$ & $3.9\times10^{-3}$ &  $1.1\times10^{-3}$ & 8.7 \\
E5 & 20 & $10^{-3}$ & $53$  & $2.5\times10^{9}$ & $1.7\times10^{3}$ & $1.7\times10^{3}$ & $4.0\times10^{-3}$ &  $1.1\times10^{-3}$ & 8.7 \\
F1 & 25 & 0         & $100$ & $4.0\times10^{9}$ & $1.1\times10^{3}$ & $1.1\times10^{3}$ & $6.9\times10^{-4}$ & -$3.6\times10^{-4}$ & 0 \\
F2 & 25 & $10^{-4}$ & $93$  & $4.0\times10^{9}$ & $1.2\times10^{3}$ & $1.2\times10^{3}$ & $1.6\times10^{-3}$ &  $3.9\times10^{-4}$ & 0.87 \\
F3 & 25 & $10^{-3}$ & $94$  & $4.0\times10^{9}$ & $1.2\times10^{3}$ & $1.2\times10^{3}$ & $1.8\times10^{-3}$ &  $5.2\times10^{-4}$ & 8.7 \\
G1 & 30 & 0         & $140$ & $5.7\times10^{9}$ & $9.7\times10^{2}$ & $9.7\times10^{2}$ & $4.5\times10^{-4}$ & $-2.5\times10^{-4}$ & 0 \\
G2 & 30 & $10^{-4}$ & $140$ & $5.7\times10^{9}$ & $9.7\times10^{2}$ & $9.7\times10^{2}$ & $4.6\times10^{-4}$ & $-2.5\times10^{-4}$ & 0.87 \\
G3 & 30 & $10^{-3}$ & $130$ & $5.7\times10^{9}$ & $1.0\times10^{3}$ & $1.0\times10^{3}$ & $9.0\times10^{-4}$ & $-3.0\times10^{-5}$ & 8.7 \\
G4 & 30 & $2\times10^{-3}$ & $110$ & $5.7\times10^{9}$ & $1.2\times10^{3}$ & $1.2\times10^{3}$ & $1.5\times10^{-3}$ &  $1.8\times10^{-4}$ & $17$ \\
G5 & 30 & 0         & $140$ & $5.7\times10^{9}$ & $9.6\times10^{2}$ & $9.6\times10^{2}$ & $4.5\times10^{-4}$ & $-2.2\times10^{-4}$ & 0 \\
\hline
\end{tabular}
\tablefoot{For each run, Pr $ = 60$,
    Pr$_{\rm M} = 1$, and Pr$_{\rm SGS} = 2.5$.}
\end{table*}

\section{Results}\label{sec:results}

We ran a total of 21 simulations separated into five sets: C,
D, E, F, and G. Each set is characterized by a fixed rotation rate of 5, 10,
20, 25, and 30 times the solar rotation rate, respectively. We varied the value
of the centrifugal force within each set.

\subsection{Differential rotation}\label{sect:diffrot}
We began by exploring changes in the differential rotation of the simulations
by defining
\begin{equation}
    \Delta_\Omega^{(60^\circ)} = \frac{\ol{\Omega}(0^\circ,s) -
    \ol{\Omega}(60^\circ, s)}{\ol{\Omega}(0,s)}\label{equ:lDR}
\end{equation}
and
\begin{equation}
        \Delta_\Omega^{(r)} = \frac{\ol{\Omega}(0^\circ,s) -
    \ol{\Omega}(0^\circ, b)}{\ol{\Omega}(0^\circ,s)}\label{equ:rDR}
\end{equation}
as measures of latitudinal and radial differential rotation. Here,
$s$ and $b$ indicate that the values are taken near the surface ($r=0.98R$) and
the bottom ($r=0.72R$),
respectively, and $\ol\Omega = \Omega_0 + \ol u_\phi/(r\sin\theta)$,
where the overbars denote azimuthal averaging, namely
\begin{equation}
  \ol{u}_\phi = \frac{1}{2\pi}\int_0^{2\pi}u_\phi(r,\theta,\phi,t)\,{\rm d}\phi.
\end{equation}
In what follows,
additional time-averaging is denoted by $\langle \cdot \rangle_t$.
Time averages of $\Delta_\Omega^{(60^\circ)}$ and
$\Delta_\Omega^{(r)}$ are listed in columns 8 and 9 of Table~\ref{tab:runs}.

\begin{figure}
\centering
\includegraphics[width=\columnwidth]{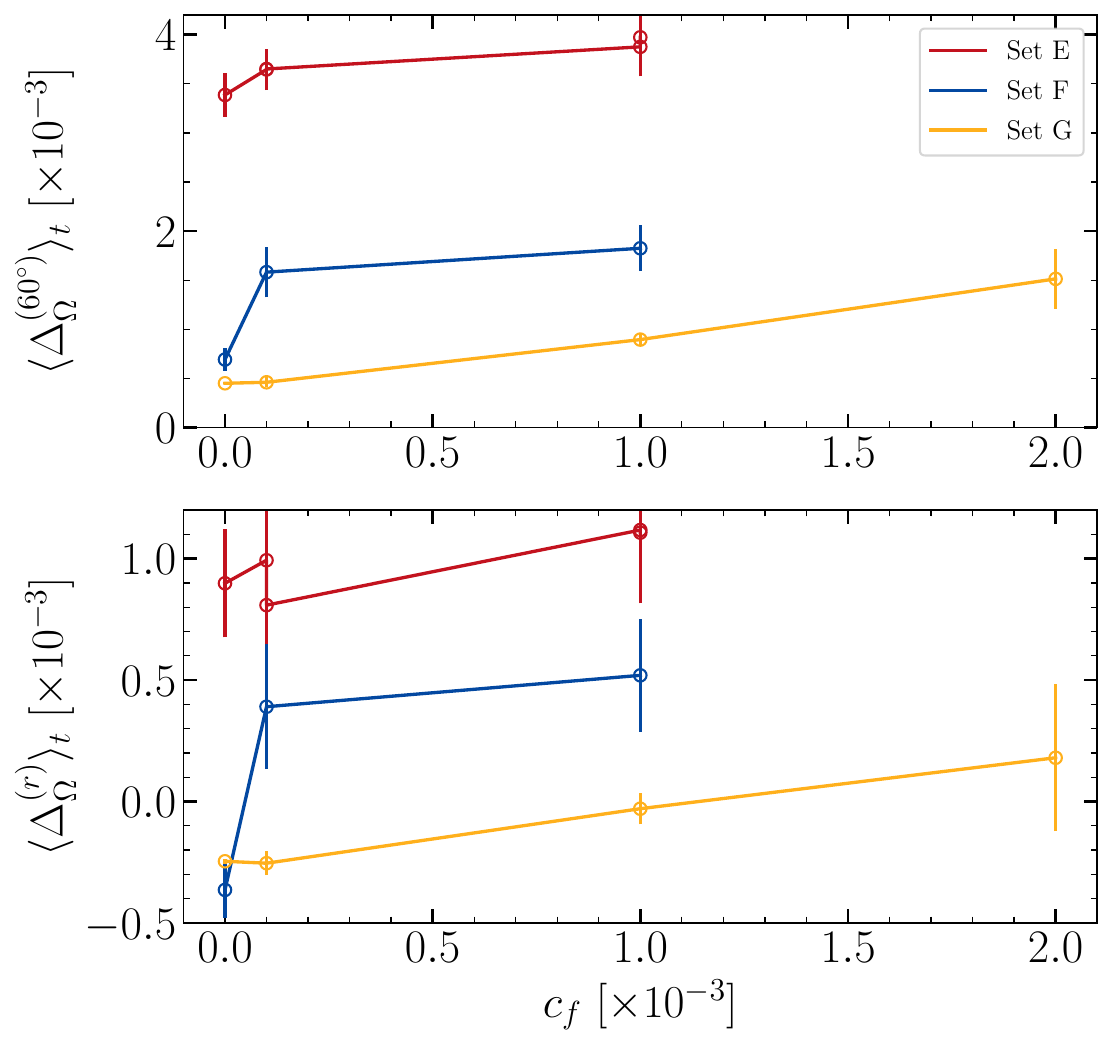}
\caption{Time-averaged differential rotation for sets E, F, and G in red, blue,
and yellow, respectively. The top and bottom panels show
$\Delta_\Omega^{(60^\circ)}$ and $\Delta_\Omega^{(r)}$ according to
Eqs.~(\ref{equ:lDR}) and (\ref{equ:rDR}), respectively.}
\label{fig:DiffRotScatter}%
\end{figure}

In the slowly rotating runs, sets C and D, there are changes in both radial and
latitudinal differential rotation, but they are very small. Comparing the 
runs without the centrifugal force with those with the largest value of $c_f$,
the biggest change in $\Delta_\Omega^{(r)}$, of about $20\%$, is in set D. However, larger deviations are found in sets E, F, and G.\ They are also
shown in Fig.~\ref{fig:DiffRotScatter} with the corresponding error bars, which
were estimated by computing the average of three equally long parts of
the time series and
taking the largest deviation from the total as the error.
\begin{figure}
\centering
\includegraphics[width=\columnwidth]{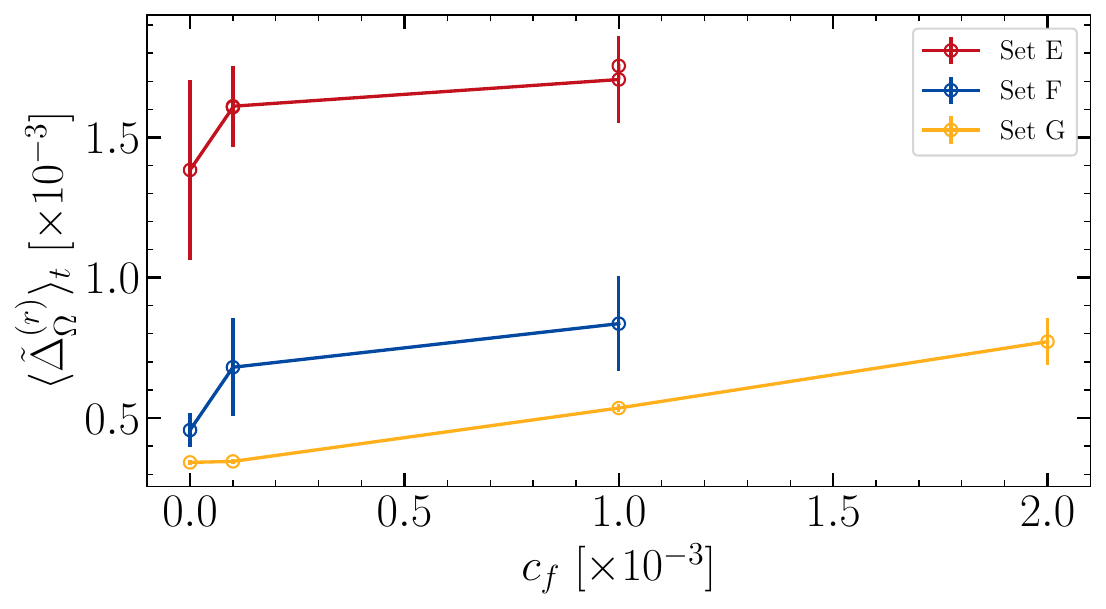}
\caption{Time-averaged radial differential rotation as defined in Eq.~(\ref{eq:rDR2}).}
\label{fig:DiffRotTilde}%
\end{figure}
In set E we see that the differential rotation of the runs that were
initialized with the same $c_f$ but at a different time, namely E2 with E3 and
E4 with E5, have very similar values. This shows that the
averaged differential rotation does not significantly depend on the
initial conditions when the centrifugal force is added. Within this
set,
the maximum deviation of $\Delta_\Omega^{(r)}$ is about $23\%$ between
runs E1 and E5. In contrast, $\Delta_\Omega^{(60^\circ)}$ is reduced
by about $17\%$.

Recently, \cite{Kapyla23} noted that the details of the radial profile
of $\ol\Omega$ can introduce spurious effects into the measure of
differential rotation as defined in Eq.~(\ref{equ:rDR}). Following their
approach, we defined the mean rotational profile at the equator as
\begin{equation}\label{eq:rDR2}
    \tilde{\Delta}^{(r)}_{\Omega} = \frac{\int_{r_{\rm in}}^{r_{\rm out}}r^2[\ol\Omega(\theta_{\rm eq},r) - 1]{\rm d}r}{\int_{r_{\rm in}}^{r_{\rm out}}r^2{\rm d}r},
\end{equation}
where $r_{\rm in} = 0.72R$ and $r_{\rm out} = 0.98R$. In Fig.~\ref{fig:DiffRotTilde}
we plot this quantity as a function of $c_f$. The differential rotation is
solar-like ($\tilde{\Delta}^{(r)}_{\Omega}>0$), as already seen in the top panel
of Fig.~\ref{fig:DiffRotScatter}. This shows that, if there are transients of
anti-solar differential rotation in our simulations, they are not very long and
a similar scaling is seen with both definitions.

As the rotation velocity increases more, the amplitude of the latitudinal
differential rotation decreases further in run F1. This is a common feature of
convection in rotating spherical shells \citep[see,
  e.g.,][]{BBBMT08,Gastine14,Viviani18}, which
is also found in Cartesian coordinates with the star-in-a-box setup
\citep{Kapyla20a}. In runs F2 and F3, $\langle\Delta_\Omega^{(60)}\rangle_t$ 
is larger by a factor of about 2.3 and 2.6.
Run F1 has bottom layers that rotate slightly faster than
the surface layers, as indicated by the negative sign of
$\langle\Delta_\Omega^{(r)}\rangle_t$. The addition of the centrifugal force
changes this pattern back to a solar-like one, where the surface layers
rotate faster, although the overall differential rotation remains weak.

We do not see major differences between runs G1 and G2, and in G3 the
latitudinal (radial) differential rotation increases (decreases) by a
factor of about 2 (10). Each of these runs has
$\langle\Delta_\Omega^{(60)}\rangle_t > 0$ and
$\langle\Delta_\Omega^{(r)}\rangle_t < 0$. In
run G4, $\langle\Delta_\Omega^{(60)}\rangle_t$ is comparable to that
in F2, and,
similarly, the radial differential rotation is shifted back to a solar-like
pattern. However, the amplitudes are all very small and close to rigid rotation.
In the control simulation (G5) we took a snapshot from run G3 and
switched off the centrifugal force; we obtained a solution that is
nearly the same as in run G1. This hints
at the possibility that the effects we are seeing are due to a systematic
effect of the centrifugal force rather than a chaotic behavior due to the
change in the initial conditions.

Overall, we find that changes in the differential rotation due to the
centrifugal force are only noticeable in the rapidly and very rapidly rotating
sets E, F, and G. We conclude that the changes in both
$\langle\Delta_\Omega^{(60)}\rangle_t$ and $\langle\Delta_\Omega^{(r)}\rangle_t$
are due to the centrifugal force and are likely insensitive to the details
of the initial conditions taken from the parent runs.
In a real star, the centrifugal force would also change the geometry of
the star. However, we cannot assess the extent of this change because
the fixed grid in our model does not allow the geometry to change.

\subsection{Magnetic energy}
\begin{figure*}
\centering
\includegraphics[width=\linewidth]{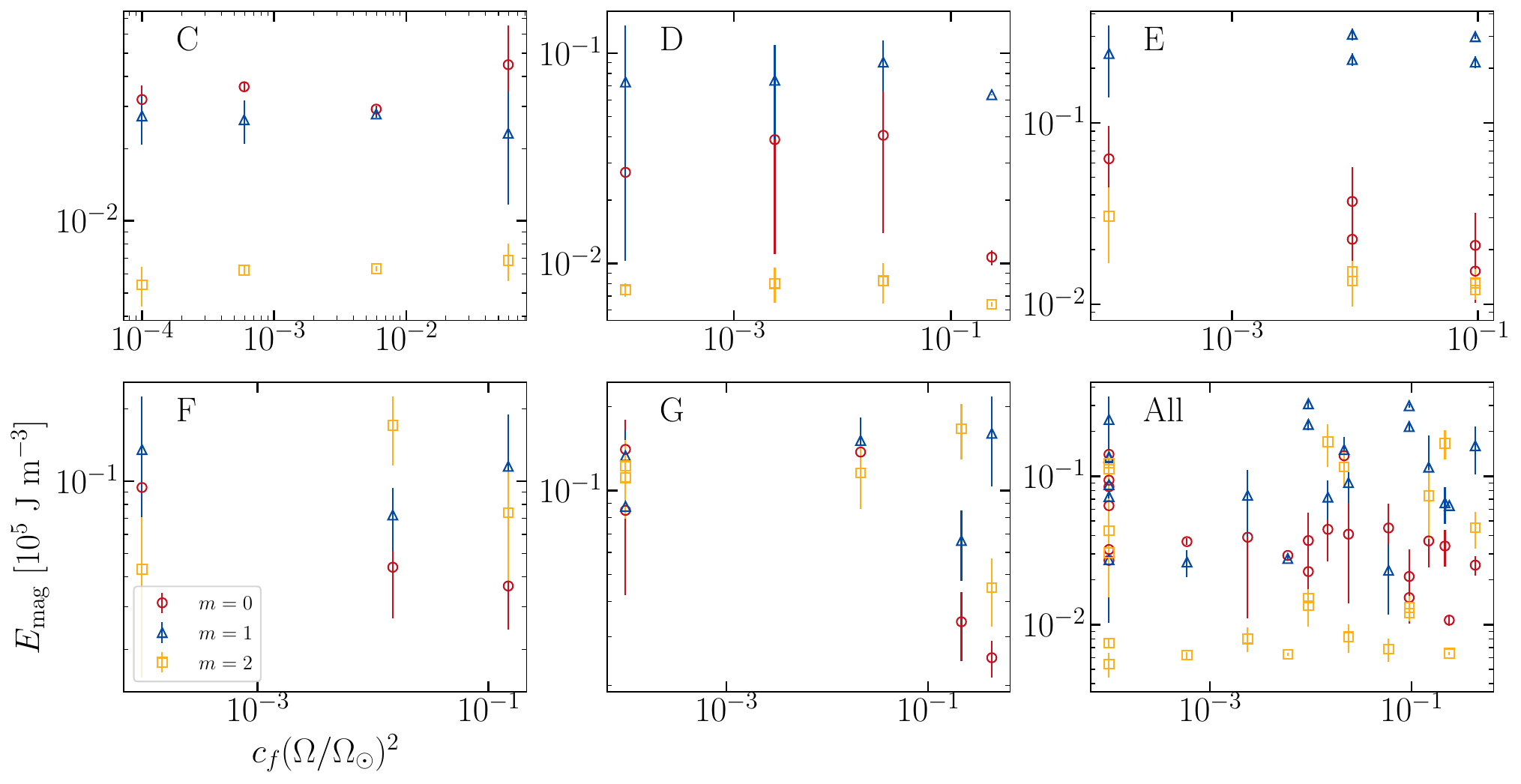}
\caption{Magnetic energy of the three lowest azimuthal modes ($m$) as a
    function
    of the centrifugal force amplitude. Runs with $c_f=0$ are given a
    fiducial value of
    $c_f(\Omega/\Omega_\odot)^2=10^{-4}$.}
\label{fig:Emag}%
\end{figure*}
\begin{table*}
\caption{Magnetic energy density from the spherical harmonic decomposition for each
run in units of $10^5$ J m$^{-3}$.}
\label{tab:energy}
\centering
\begin{tabular}{c c c c c c}
  \hline\hline
  Run & $E_{\rm mag}^{m=0}$ & $E_{\rm mag}^{m=1}$ & $E_{\rm mag}^{m=2}$ \\
  \hline
  C1 & $3.2\times10^{-2} \pm 4.7\times10^{-3}$ & $2.7\times10^{-2} \pm 6.6\times10^{-3}$ & $5.4\times10^{-3} \pm 1.0\times10^{-3}$ \\
  C2 & $3.6\times10^{-2} \pm 1.8\times10^{-3}$ & $2.6\times10^{-2} \pm 5.4\times10^{-3}$ & $6.2\times10^{-3} \pm 3.3\times10^{-4}$ \\
  C3 & $2.9\times10^{-2} \pm 1.0\times10^{-3}$ & $2.8\times10^{-2} \pm 1.0\times10^{-3}$ & $6.3\times10^{-3} \pm 1.5\times10^{-4}$ \\
  C4 & $4.5\times10^{-2} \pm 2.1\times10^{-2}$ & $2.3\times10^{-2} \pm 1.1\times10^{-2}$ & $6.8\times10^{-3} \pm 1.2\times10^{-3}$ \\
  D1 & $2.7\times10^{-2} \pm 1.4\times10^{-2}$ & $7.3\times10^{-2} \pm 6.3\times10^{-2}$ & $7.5\times10^{-3} \pm 5.4\times10^{-4}$ \\
  D2 & $3.9\times10^{-2} \pm 2.8\times10^{-2}$ & $7.4\times10^{-2} \pm 3.5\times10^{-2}$ & $8.0\times10^{-3} \pm 1.5\times10^{-3}$ \\
  D3 & $4.1\times10^{-2} \pm 2.7\times10^{-2}$ & $9.1\times10^{-2} \pm 2.5\times10^{-2}$ & $8.3\times10^{-3} \pm 1.8\times10^{-3}$ \\
  D4 & $1.1\times10^{-2} \pm 8.8\times10^{-4}$ & $6.3\times10^{-2} \pm 2.0\times10^{-4}$ & $6.4\times10^{-3} \pm 1.5\times10^{-4}$ \\
  E1 & $6.3\times10^{-2} \pm 3.3\times10^{-2}$ & $2.4\times10^{-1} \pm 1.0\times10^{-1}$ & $3.0\times10^{-2} \pm 1.4\times10^{-2}$ \\
  E2 & $2.3\times10^{-2} \pm 1.3\times10^{-2}$ & $2.2\times10^{-1} \pm 1.8\times10^{-2}$ & $1.3\times10^{-2} \pm 3.7\times10^{-3}$ \\
  E3 & $3.7\times10^{-2} \pm 2.0\times10^{-2}$ & $3.1\times10^{-1} \pm 2.4\times10^{-2}$ & $1.5\times10^{-2} \pm 2.2\times10^{-3}$ \\
  E4 & $2.1\times10^{-2} \pm 1.1\times10^{-2}$ & $2.2\times10^{-1} \pm 1.7\times10^{-2}$ & $1.3\times10^{-2} \pm 2.5\times10^{-3}$ \\
  E5 & $1.5\times10^{-2} \pm 3.9\times10^{-3}$ & $3.0\times10^{-1} \pm 7.9\times10^{-3}$ & $1.2\times10^{-2} \pm 4.0\times10^{-5}$ \\
  F1 & $9.4\times10^{-2} \pm 7.8\times10^{-2}$ & $1.4\times10^{-1} \pm 8.9\times10^{-2}$ & $4.3\times10^{-2} \pm 2.8\times10^{-2}$ \\
  F2 & $4.4\times10^{-2} \pm 1.7\times10^{-2}$ & $7.2\times10^{-2} \pm 2.1\times10^{-2}$ & $1.7\times10^{-1} \pm 5.4\times10^{-2}$ \\
  F3 & $3.7\times10^{-2} \pm 1.2\times10^{-2}$ & $1.2\times10^{-1} \pm 7.4\times10^{-2}$ & $7.4\times10^{-2} \pm 3.6\times10^{-2}$ \\
  G1 & $1.4\times10^{-1} \pm 3.9\times10^{-2}$ & $1.3\times10^{-1} \pm 3.2\times10^{-2}$ & $1.2\times10^{-1} \pm 3.0\times10^{-2}$ \\
  G2 & $1.4\times10^{-1} \pm 4.3\times10^{-2}$ & $1.5\times10^{-1} \pm 3.2\times10^{-2}$ & $1.2\times10^{-1} \pm 3.0\times10^{-2}$ \\
  G3 & $3.4\times10^{-2} \pm 9.4\times10^{-3}$ & $6.6\times10^{-2} \pm 1.9\times10^{-2}$ & $1.7\times10^{-1} \pm 3.7\times10^{-2}$ \\
  G4 & $2.5\times10^{-2} \pm 3.8\times10^{-3}$ & $1.6\times10^{-1} \pm 5.7\times10^{-2}$ & $4.5\times10^{-2} \pm 1.2\times10^{-2}$ \\
  G5 & $8.5\times10^{-2} \pm 4.3\times10^{-2}$ & $8.8\times10^{-2} \pm 4.6\times10^{-3}$ & $1.1\times10^{-1} \pm 3.2\times10^{-2}$ \\
  \hline
\end{tabular}
\end{table*}

The magnetic energy of the first three azimuthal modes near the surface are listed in
Table~\ref{tab:energy} and shown as a function of the centrifugal force
amplitude in Fig.~\ref{fig:Emag}. It is defined as
\begin{equation}
  E_{\rm mag}^{m=i} = \frac{1}{2\mu_0}\left\langle\sum_{l\geq m} {{\bm B} ^2_{l,m=i}}\right\rangle_{\theta\phi t},
\end{equation}
where $B_{l,m=i}$ are obtained from the spherical harmonic decomposition.
At slow rotation, sets C and D do not
show significant changes in the energy as the centrifugal force increases.
At the same time, we also see that in set C the axisymmetric mode dominates the
runs. This contrasts with the previous study of \cite{Viviani18}, who find that, at rotation rates larger than $\Omega/\Omega_\odot\sim1.8$,
the $m=1$ mode dominates the runs, and after $\Omega/\Omega_\odot\sim20$
the dominance falls back to $m=0$. However, at higher grid resolutions, they
find that this trend is suppressed and so the $m=1$ mode
dominated again. In the current simulations, we find that this trend
only starts to show up in set D. In all cases, the $m=2$ mode is always
subdominant by a factor of roughly 10.

Similarly to the differential rotation,
the effects of the centrifugal force are more noticeable in sets E, F, and G.
In this rapidly rotating regime, the axisymmetric mode is always subdominant.
In set E, the $m=1$ mode always has the highest energy, and as the amplitude
of the centrifugal force increases, $E_{\rm mag}^{m=0}$ decreases and so does
$E_{\rm mag}^{m=2}$. We do not see noticeable differences between runs E2 and
E3, which have the same centrifugal force but were initialized at different
times. This is also the case for runs E4 and E5, meaning no hysteresis is observed and the results are independent of the
history of the run. In set F, we see that the
energy in the $m=0$ mode decreases, and in run F2 the $m=2$
mode carries most of the energy. However, when the centrifugal force is
increased further, the $m=1$ mode becomes dominant once again.

In run G1 there is no clearly dominating mode, and the energy in the $m=0$
mode is only roughly $5\%$ larger than in $m=1$. As the centrifugal force is first
added in run G2, the energy
of the $m=0$ mode increases by about $10\%$. However, similarly to run F2, run G3
has most of the magnetic energy in the $m=2$ mode, which is about
$66\%$ higher than $E_{\rm mag}^{m=0}$ and $E_{\rm mag}^{m=1}$ combined.
Increasing the centrifugal force further, we see that the dominant mode
is $m=1$, same as in the case of run F3. When the centrifugal force is switched
off, the distribution of the energy goes back to levels nearer to run
G1 with $c_f=0$.

\subsection{Azimuthal dynamo waves}\label{sec:adw}
We began by estimating the period of the ADWs by building a periodogram
and identifying the signal with the greatest power as the main cycle.
We then investigated whether there are tendencies between the period of
the ADW and the rotation rate in runs without the centrifugal force.
\begin{figure}
\centering
\includegraphics[width=\columnwidth]{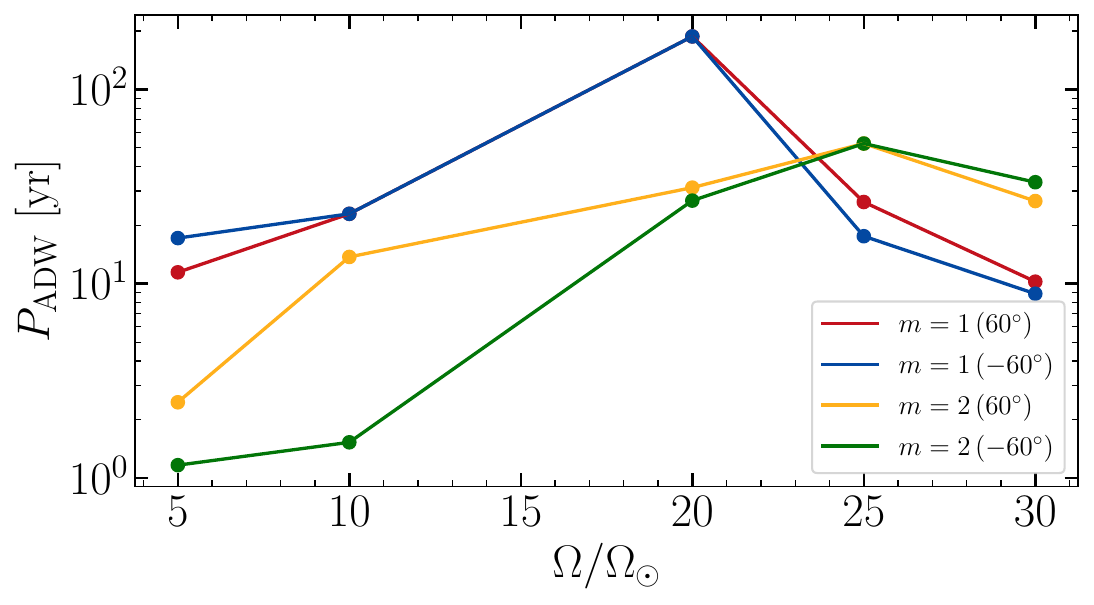}
\caption{Period of the ADWs as a function of
  the normalized rotation rate
  for runs without the centrifugal force.}
\label{fig:PawdOmega}%
\end{figure}
This is shown in Fig.~\ref{fig:PawdOmega}. From $\Omega~=~5\Omega_\odot$ to
$\Omega~=~20\Omega_\odot$, the period of the ADWs of the $m=1$ and
$m=2$ modes seems to increase with rotation. For more rapid rotation,
the period of the $m=1$ ADW decreases, whereas for the $m=2$ mode this
tendency appears for $\Omega \geq 25\Omega_\odot$. For the two most
rapidly rotating cases, the period of the $m=2$ ADW exceeds that of the
$m=1$ mode.

\begin{figure*}
\centering
\includegraphics[width=\textwidth]{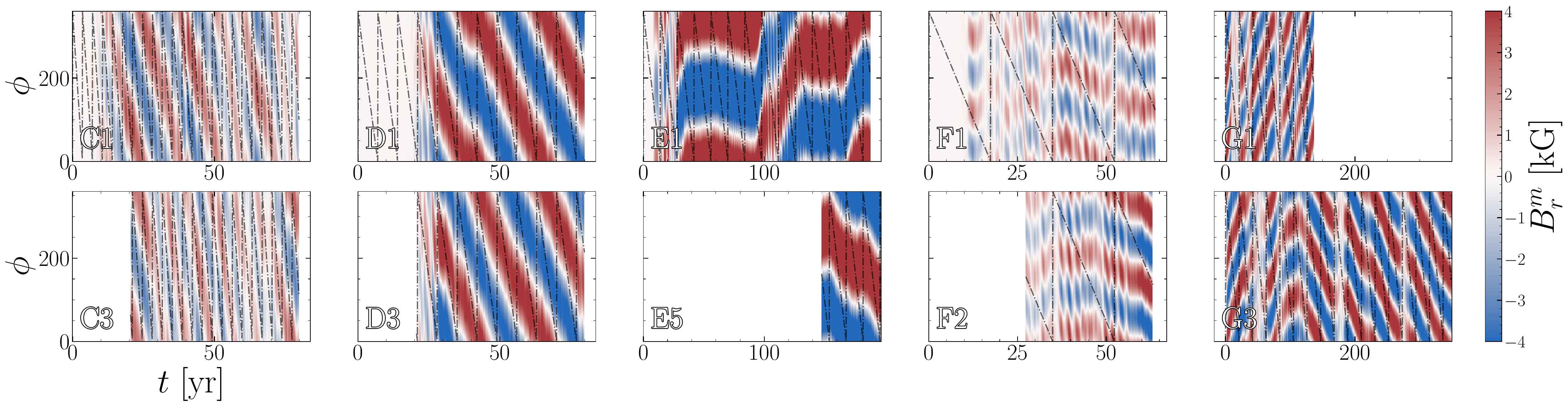}
\caption{ADWs for some selected runs at a latitude of
$\theta = 60^\circ$. The top panels are the runs without the centrifugal force.
The dashed line denotes the path that the ADWs would follow if they were
advected by the differential rotation. For each set, we have added null data
to make the time axes coincide in order to facilitate comparison.}
\label{fig:ADW2}%
\end{figure*}

\begin{table}
    \caption{Properties of the ADWs.}
\label{tab:ADWproperties}
\centering
\begin{tabular}{c c c}
  \hline\hline
  Run & $P_{\rm ADW}$ [yr] & Propagation \\
  \hline
  C1 & 11.44$_{m_1}^*$           &  R     \\
  C2 & 9.64$_{m_1}^*$            &  R     \\
  C3 & 8.48$_{m_1}$              &  R     \\
  C4 & 8.80$_{m_1}$              &  R     \\
  D1 & 22.88$_{m_1}$             &  R     \\
  D2 & 22.45$_{m_1}$             &  R     \\
  D3 & 19.73$_{m_1}$             &  R     \\
  D4 & 21.60$_{m_1}$             &  R     \\
  E1 & >187.30$_{m_1}$           &  S,P   \\
  E2 & -                         &  S     \\
  E3 & >116.40$_{m_1}$           &  R,P \\
  E4 & >64.07$_{m_1}$            &  R     \\
  E5 & >73.48$_{m_1}$            &  R     \\
  F1 & >52.64$_{m_2}^\dagger$    &  R,P   \\
  F2 & >35.84$_{m_2}$            &  R     \\
  F3 & >15.04$_{m_2}$            &  R   \\
  G1 & 26.62$_{m_2}^{*,\dagger}$ &  P     \\
  G2 & 40.74$_{m_2}^{*,\dagger}$ &  P     \\
  G3 & 69.68$_{m_2}$             &  R     \\
  G4 & 31.65$_{m_2}^*$           &  R     \\
  G5 & 65.60$_{m_2}^*$           &  P     \\
  \hline
  \hline
  \end{tabular}
\tablefoot{Data were taken at latitude
  $\theta = 60^\circ$ for each run.
  The ``greater than'' symbol indicates that the period of the
  ADW is not covered in the simulated time.
  Asterisks denote a difference between the
  period of the ADW at the opposite latitude, and ``$\dagger''$ denotes nearly
  equally strong $m=1$ and $m=2$ signals. S, R, and P stand for
  standing, retrograde,
  and prograde propagation. In the case of Run E1, a wave with a period of
  about 80 years can also be identified.}
\end{table}

To explore the migration pattern of the ADW, we show in Fig.~\ref{fig:ADW2}
the $m=1$ mode of the radial magnetic field near the surface at a latitude of
$60^\circ$ for runs C1, C3, D1, D3, E1, and E5, as well as the $m=2$ mode for
runs F1, F2, G1, and G3. Overplotted is the advection path due to differential
rotation. In Table~\ref{tab:ADWproperties} we list the periods of the ADWs at
$\theta=60^\circ$ and the direction of the propagation.

In runs C1 and C3 we obtain a retrograde migration with no evidence of changes
due to the centrifugal force. Both migration patterns appear to be constant in
time with no interruptions. Similarly, the ADWs in runs D1 and D3
have a retrograde migration pattern but are characterized by a longer
period. Run E1 has an interesting non-axisymmetric dynamo solution
that shows periods of prograde and standing ADWs for the $m=1$
mode. The migration of the ADW is changed by the centrifugal force, as
evidenced by the
panel for run E5. This run has a retrograde migration, similarly to
sets C and D, and shows no similarity to run E1. As shown in
Sect.~\ref{sect:diffrot}, the change in the latitudinal differential rotation
between runs E1 and E5 is about 17\%. However, the ADWs propagate
almost like rigid structures, so differential rotation cannot directly
be used to explain their behavior. The precise origin of ADWs is
unclear even in the case where the centrifugal force is absent, but quantities relevant for large-scale dynamos, such as
differential rotation, kinetic helicity, and other turbulent
quantities, along with their spatio-temporal profiles, likely play
roles. However, the changes we observe when the centrifugal force is
included suggest that subtle changes in the velocity field are
enough to significantly alter the behavior of ADWs.

\begin{figure}
\centering
\includegraphics[width=\columnwidth]{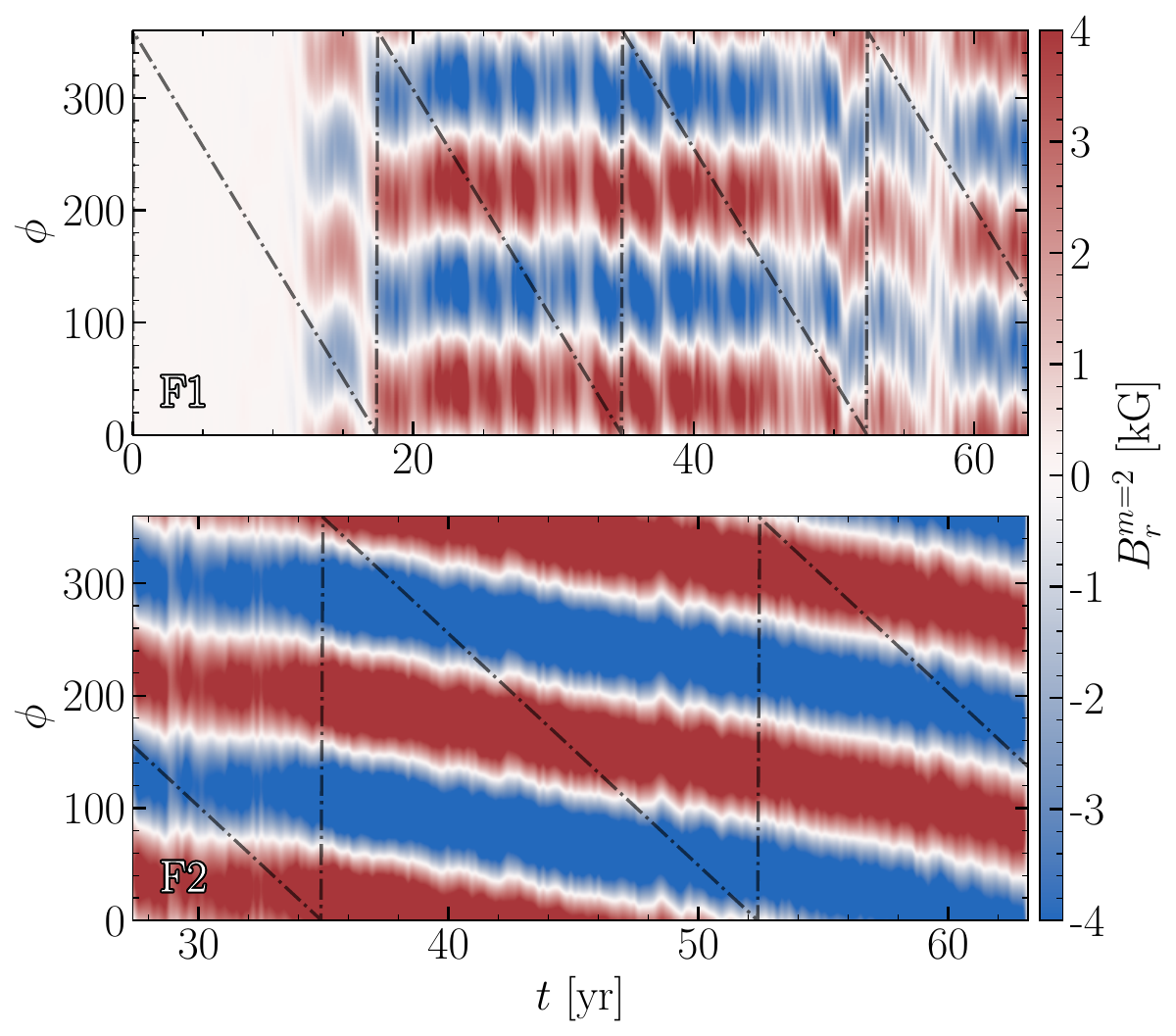}

\caption{Same as Fig.~\ref{fig:ADW2} but at $\theta = -60^\circ$ for runs F1
(top) and F2 (bottom).}
\label{fig:F1F2m2}%
\end{figure}
In run F1 the wave is standing or very slowly migrating in a retrograde
direction. The migration period seems to decrease as the centrifugal force
is added in the lower panel of Fig.~\ref{fig:F1F2m2}, where the wave travels
about $120^\circ$ in azimuth. Also evident here is
the increase in the magnitude of the $m=2$ mode at the southern hemisphere.
It also seems that this part contributes the most to the change in magnetic
energy seen in Fig.~\ref{fig:Emag}. It is only toward the end of the simulations that the
$B_r^{m=2}$ at the northern hemisphere catches up and becomes comparable
in strength to the southern hemisphere counterpart, as can be seen by
comparing the panels of run F2 in Figs.~\ref{fig:ADW2} and
\ref{fig:F1F2m2}.

In the rapidly rotating regime, the $m=2$ mode of run G1 has a
periodic wave with a period of about 26 years (see Table~\ref{tab:ADWproperties}),
with clear prograde propagation. The centrifugal force changes
the propagation direction, as can be seen in the last panel of
Fig.~\ref{fig:ADW2} (run G3), and the period of the ADW is also affected
such that now it is about 70 years. In this case, the latitudinal differential
rotation is doubled in run G3 as compared to run G1. Although this is
the most obvious change between the simulations, it is difficult to
explain the change in the ADWs with this alone, as discussed above.

In general, we find a preference for retrograde propagation, as 
in \cite{Viviani18}. Interestingly, however, run E1 shows
a combination of standing and prograde waves and G1 is prograde.
A subsequent study by \cite{Viviani21}, where the prescribed heat conductivity
was replaced by the Kramers opacity law, showed that there is a tendency of
producing prograde-propagating ADWs. We replaced the fixed radial
profile $K(r)$ with the corresponding quantity from Kramers opacity
(see Eq.~\ref{eq:Kramers}) and branched run E1 off to a new run, K1.
The centrifugal force was added in run K2 with $c_f=5\times10^{-4}$.
\begin{figure}
\centering
\includegraphics[width=\columnwidth]{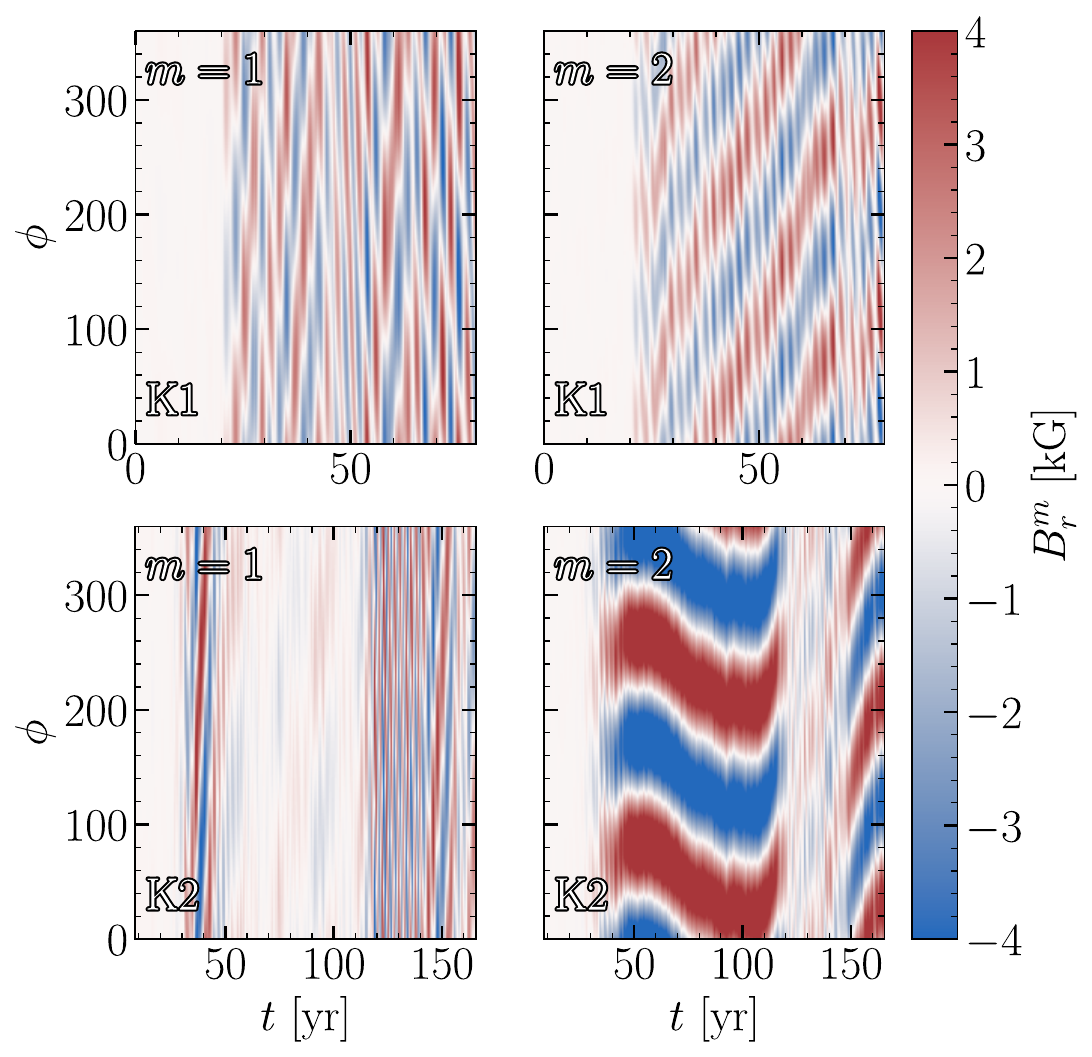}
\caption{ADWs for the runs with Kramers opacity without the centrifugal force
(top) and with the centrifugal force (bottom).}
\label{fig:K1K2}%
\end{figure}
Figure~\ref{fig:K1K2} shows the reconstructed $m=1,2$ modes for these two runs.
It is clear that run E1 is different than K1, as the properties of the ADW are
not reproduced when the Kramers opacity is used. In the latter case, the ADW
is prograde for both $m=1$ and $m=2$ modes, which also have comparable
energies.
This is in accordance to \cite{Viviani21} in that it seems as if prograde
migration is favored when the Kramers opacity is used. When the centrifugal
force is added in run K2, the strength of the first non-axisymmetric mode
decreases down to around $1$~kG, but the direction of the propagation appears
to be unaffected.
Interestingly, the $m=2$ mode increases from about $4$~kG in run K1 to
roughly $6$~kG
in run K2. The propagation pattern is interesting in that it seems to oscillate
around a mean azimuth with an amplitude of $40^\circ$ between $t=40$~yr and
$t=110$~yr. After this, the strength of the $m=2$ ($m=1$) mode decreases 
(increases), and at around $t=150$~yr the $m=2$ mode reappears without a
corresponding decrease in the $m=1$ mode.
In contrast to set E, the combination of Kramers opacity and the centrifugal
force produces a dominant $m=2$ mode, which was only found in the more
rapidly rotating runs F2 and G3.

\section{Summary and conclusions}\label{sec:conclusions}
In this paper we have studied the effects of the centrifugal force in
semi-global dynamo simulations. It is important to assess its effects in the
context of young solar analogs, which are used to study the Sun in an
astrophysical context.

The amplitude of the centrifugal force is considered in our setup as a
free parameter and is
thus decoupled from the Coriolis term and the rotation of the star. In this
way, its amplitude is artificially reduced by the control parameter
$c_f$. This allows us to avoid an unrealistically large centrifugal
force
\citep{Kapyla13}. This approach was applied to a total of 21 simulations
divided into five sets, each characterized by a different rotation rate and with
different values of $c_f$ within each set. We find that the centrifugal force
induces changes in the differential rotation and magnetic field only when
rotation is rapid enough.

Both the latitudinal and radial differential rotation tend to increase with
increasing centrifugal force. In the two most rapidly rotating runs without
the centrifugal force, we obtained an anti-solar radial differential rotation.
After including the centrifugal force, this solutions changed to solar-like
differential rotation. Namely, the outer layers of the convection zone went
from rotating more slowly to more rapidly relative to the deeper layers.
All of our runs have a solar-like latitudinal differential rotation, where
high latitudes rotate more slowly than the regions nearer to the equator. This
difference increases as the centrifugal force becomes stronger
(see Fig.~\ref{fig:DiffRotScatter}).

The magnetic energy, shown in Fig.~\ref{fig:Emag}, also shows noticeable effects
only when the rotation is rapid enough in sets E, F, and G. All runs
are dominated by the $m=0$ mode in set~C and by the $m=1$ mode in set~D, and they show small
changes in energy as a function of the centrifugal force amplitude.
In the rapidly rotating regime, it is common to find a dominating $m=1$ mode,
and the energy of the $m=0$ and $m=2$ modes decreases as $c_f$ increases in
set~E. This trend is also present in sets F and G, with the difference that
there are some cases where the $m=2$ mode dominates.

By analyzing the ADWs near the surface of our runs, we find that the direction
of the propagation changes from prograde to retrograde in some rapidly rotating
runs as a function of the centrifugal force. This is most easily seen in runs
E5 and G3 in Fig.~\ref{fig:ADW2}. For run F2, we find that the direction of the
propagation of the ADW is not clearly affected, but there are indications that
its period might be affected (see Fig.~\ref{fig:F1F2m2}).

To confirm the effects of the centrifugal force, we introduced three control
runs. First, in order to test the importance of the initial conditions, we started run E3 (E5) with the same value of $c_f$ as E2 (E4) from run
E1 but at a different time. There are negligible differences in the differential
rotation, as can been seen in columns 8 and 9 of Table~\ref{tab:runs} and
in the overlapping points at constant $c_f$ in Fig.~\ref{fig:DiffRotScatter}.
The magnetic energy is only slightly affected, as seen in the third panel of
Fig.~\ref{fig:Emag} from the data points at constant
$c_f(\Omega/\Omega_\odot)^2$. Secondly, it is important to look at the solution
of a run with the centrifugal force when it is turned off again. We did this
experiment with run G3, in which the propagation of the ADW was retrograde
(see the last panel of Fig.~\ref{fig:ADW2}).
When the centrifugal force was turned off, the propagation changed back to
prograde, as it was in the original run, G1. Overall, the control runs show
that the changes described above are due to the centrifugal force and not
likely the outcome of the nonlinear evolution of the equations.

A previous study by \cite{Viviani21} shows that the propagation of the
ADWs can be affected by the introduction of the Kramers opacity instead
of a fixed radial profile of heat conductivity. We combined the
Kramers opacity with the
centrifugal force in runs K1 and K2 and find that, first, the solution
of the ADW is different for the runs without the centrifugal force (K1) and
with the centrifugal force (K2) as compared with the corresponding runs with the
spatially fixed heat conductivity (set E). Notably, run K2 showed a
migration pattern that was
not obtained in any of the other runs (see Fig~\ref{fig:K1K2}). This confirms
that the Kramers opacity changes the ADW solution, but it is even more complex
when the centrifugal force is included.

Despite our experiments, we were unable to identify
the mechanism responsible for changing the behavior of the
ADWs. The clearest change due to the centrifugal force is seen in the
differential rotation, but its effect must be indirect through the
dynamo mechanism because advection by a shear flow is incompatible
with the practically rigidly propagating ADWs. The details of the
dynamo process in 3D simulations are highly complex
\citep[e.g.,][]{2021ApJ...919L..13W}, and current mean-field methods are
applicable only in the axisymmetric case. Observations, \rm{specifically}
an analysis of the surface
magnetic field and its cycles as a function of rotation, could help us better understand this.
Such a study was
performed by \cite{Lehtinen16}, who find that the photometric rotation period and
activity period of a group of stars
show clear differences. They proposed that this trend could be
explained by the presence of prograde ADWs. However, our simulations
suggest that when a prograde wave is affected by the centrifugal force,
it changes to a retrograde propagation. Such a discrepancy could be
better understood by extending the observations and by performing more
realistic
simulations in a wider parameter regime.

\begin{acknowledgements}
We thank the anonymous referee for their comments.
FHN acknowledges funding from the Deutscher Akademischer Austauschdienst.
FHN and PJK would like to thank the Isaac Newton Institute for Mathematical
Sciences, Cambridge, for support and hospitality during the programme
``Frontiers in dynamo theory: from the Earth to the stars'' where work on this
paper was undertaken. This work was supported by EPSRC grant no EP/R014604/1.
PJK acknowledges the support from the Deutsche Forschungsgemeinschaft
Heisenberg programme (grant No.\ KA 4825/4-1). PJK was also partially
supported by a grant from the Simons Foundation.
DRGS thanks for funding via the  Alexander von Humboldt Foundation, Bonn, Germany.
RB acknowledges support by the Deutsche Forschungsgemeinschaft (DFG, German
Research Foundation) under Germany’s Excellence Strategy – EXC 2121
``Quantum Universe'' – 390833306.
The authors gratefully acknowledge the computing time granted by the Resource
Allocation Board and provided on the supercomputer Lise and Emmy at NHR@ZIB
and NHR@G\"ottingen as part of the NHR infrastructure. The calculations for
this research were conducted with computing resources under the project
hhp00052.
\end{acknowledgements}

\bibliographystyle{aa} 
\bibliography{bibliography} 

\end{document}